# Secondary Peak on Asymmetric Magnetization Loop of Type-II Superconductors

D.M. Gokhfeld

L.V. Kirensky Institute of Physics SB RAS, Krasnoyarsk, 660036

**Abstract** Asymmetric magnetization loops with a second peak effect were parameterized by the extended critical state model. The magnetic field distribution in a sample is considered. Expression is suggested for a peak of the critical current density and corresponding depression on field dependence of the depth of surface layer with equilibrium magnetization. These functions determine the width and the asymmetry of a magnetization loop. Asymmetry of the secondary peak height on magnetization branches for increasing and decreasing field is reproduced on the computed magnetization curves.

## 1 Introduction

A secondary peak (fish tail) on field dependence of magnetization is observed on several bulk superconducting materials ([1] and references therein). Underlying increase of pinning at high fields is caused by transition of the vortex lattice [2,3]. Phase separation at high fields also can produce the fishtail effect [4,5].

In work [1] the critical state model (CSM) capable of describing the fishtail effect was introduced. CSM relates the magnetization with the critical current such that a secondary peak on magnetization loop is resulted from a peak on field dependence of the critical current density $j_c(B)$. This $j_c$ peak can be expressed by a function adding to an usual decreasing $j'_c(B)$ dependence (for instance the Kim model or the exponential model dependence):

$$j_c(B) = j'_c(B) + a\, j_{c0}\, f_{\text{peak}}(B), \qquad (1)$$

where $j_{c0}$ is the critical current density at $B = 0$, $a$ is the height of the peak with respect to $j_{c0}$, $f_{\text{peak}}(B)$ is a function equal to 1 at the peak position, $B_{\text{peak}}$. The Lorentzian [1] or Gaussian [6] terms were found to be appropriate for $f_{\text{peak}}(B)$. This approach describes symmetric hysteretic $M(H)$ dependencies of hard superconductors and the secondary peak. However the condition $j_c(B) = j_{c0}$ at $B = 0$ may be violated at low and medium fields. Furthermore simple CSM fails to fit the magnetization loops which have asymmetry relative to the $M = 0$ axis.

Recent the extended critical state model (ECSM) was developed and used to treat asymmetric magnetization loops [7,8]. ECSM considers the magnetization of superconductor as sum of equilibrium and nonequilibrium (irreversible) contributions. This paper is devoted to application of ECSM for both symmetric and asymmetric magnetization loops with the secondary peak. The field dependence of the critical current density is suggested in a new form to fit experimental data in a consistent manner.

## 2 Magnetization of type-II superconductors

Magnetization of superconductor is given by

$$\mu_0 M(H) = -\mu_0 H + \frac{2}{R^2}\int_0^R rB(r)dr, \qquad (2)$$

where $2R$ is the sample size in the plane perpendicular to $H$, $B(r)$ is the magnetic induction in the sample. Experimental $M$-$H$ loops are fitted by assigning the appropriate distribution of the magnetic field in the sample and the field dependence of the critical current density. The $B(r)$ dependence is found from the critical state equation:

$$dB(r)/dr = \pm \mu_0 j_c(B). \qquad (3)$$

ECSM supposes that the equilibrium magnetization realizes in the layer at the sample surface [7,8]. Screening currents act on vortices [9] and decrease pinning in this region. In work [8] the corresponding $B(r)$ distributions were obtained. Separate sets of equations (4a,b,c) are resulted for three main parts of the $M(H)$ hysteresis. During the initial field increasing from 0 to maximal value $H_m$, the $B(r)$ dependence is determined by equation:

$$\Phi(B) - \Phi(\mu_0 H) = -\mu_0 j_{c0}(R-r), \qquad (4a)$$

where $\Phi(B)$ is a function, such that $d\Phi(B)/dB = (j_c(B)/j_{c0})^{-1}$ [10]. For decreasing $H$ from $H_m$ to 0 ($M^+(H)$ branch), the $B(r)$ dependence consists of three parts which are described by equations:

$$\Phi(B) - \Phi(\mu_0 H) = -\mu_0 j_{c0}(R-r),$$

$$\Phi(B) - \Phi(B_s) = \mu_0 j_{c0}(R-r-L_s),$$

$$\Phi(B) - \Phi(\mu_0 H_m) = -\mu_0 j_{c0}(R-r), \qquad (4b)$$

where $L_s$ is the depth of the layer with equilibrium magnetization, $B_s$ is the induction value at $r = R-L_s$. Then, for changing $H$ from 0 to $-H_m$ ($M^-(H)$ branch), also three curves are resulted:

$\Phi(-B) - \Phi(-\mu_0 H) = -\mu_0 j_{c0} (R-r),$

$\Phi(B) + \Phi(-\mu_0 H) = \mu_0 j_{c0} (R-r),$

$\Phi(B) - \Phi(\mu_0 H_m) = -\mu_0 j_{c0} (R-r).$  (4c)

Asymmetry of a magnetization loop is determined by reciprocal positions of the $M^-(H)$ and $M^+(H)$ branches. Since the $M^+(H)$ branch position is affected by $L_s$ (4b), the asymmetry correlates with the parameter $P_a = L_s/R$ such that $0 \leq P_a \leq 1$. The value of $L_s$ is determined by pinning and screening capability [9] such that $L_s(B)$ is a increasing function of the magnetic field at the sample surface. While the fishtail effect is absent this field dependence of the depth of the layer with equilibrium magnetization is supposed to be simple:

$$L'_s(B) = L_{s0}\left[1 + \frac{|B|}{B^*}\right], \quad (5)$$

here $L_{s0}$ is the $L'_s$ value at $H = 0$, which is about the London penetration depth [8], parameter $B^*$ sets an increase of $L'_s$.

To introduce the magnetization peak we should trace the $j_c(B)$ and $L_s(B)$ behavior during pinning increase. Increase of $j_c$ improves the screening that should form a depression on the $L_s(B)$ dependence. This depression and the peak on the $j_c(B)$ dependence can be expressed by the single empirical function:

$$f_{peak}(B) = \exp\left(-\frac{\left(\ln\frac{|B|}{B_{peak}}\right)^2}{2\left(\frac{B_w}{B_{peak}}\right)^2}\right), \quad (6)$$

where the peak is described by its center position $B_{peak}$, the width $B_w$ and the height $A$ related to the value of $j'_c(B)$ at $B = B_{peak}$. Here the multiplier $B/B_{peak}$ is used to satisfy the conditions $j_c(B) = j_{c0}$ and $L_s(B) = L_{s0}$ at $B = 0$. Then we reorganized equation (1) such that the peak height is measured from the $j'_c(B)$ curve:

$$j_c(B) = j'_c(B)\left(1 + A f_{peak}(B)\right), \quad (7)$$

where $A = a j_{c0} / j'_c(B)$ is the peak height related to the value of $j'_c(B)$ at $B = B_{peak}$. Correspondingly

$$L_s(B) = \frac{L'_s(B)}{1 + A f_{peak}(B)}. \quad (8)$$

## 3 Discussion

Various magnetization loops with second peaks are reproduced by approach described in Section 2. Figure 1a demonstrates the magnetization loops with different asymmetry related to the parameter $P_{a0} = L_{s0}/R$. For $P_a = 0$ the magnetization loop is symmetric that is realized for hard superconductors. While pinning decreases, $P_a$ increases and the $M^+(H)$ branch is situated closer to the $M^-(H)$ curve. Distance between the branches decreases fast, as $P_a$ increases. For $P_{a0} = 0.3$ the $M^+(H)$ branch is near the $M^-(H)$ one. Given $P_a = 1$, pinning is absent and the $M^-(H)$ and $M^+(H)$ branches coincide such that the $M(H)$ dependence is reversible. Figure 1b shows evolution of the $M(H)$ dependence as the peak height $A$ increases from 0 to 1.5.

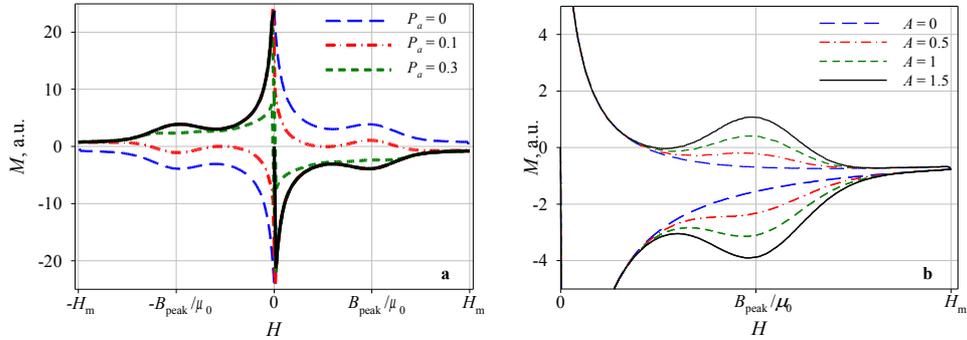

**Fig. 1** Magnetization loops of II-type superconductor. (**a**) $M(H)$ dependencies with $A = 1.5$ and different asymmetry $P_{a0} = 0, 0.1, 0.3$. Solid line is for $M^-(H)$ branch which does not depend on $P_{a0}$. (**b**) $M(H)$ dependencies with $P_{a0} = 0.1$ and different secondary peak height $A = 0, 0.5, 1, 1.5$.

In the presented model there is no discrepancy of the center peak positions on the $M^-(H)$ and $M^+(H)$ branches that disagrees with work of Tulapurkar [11]. History effect and metastability are not considered above but they may be accounted by the use of the history dependent critical current density [12,13].

Both vortex lattice transition and phase separation may produce a non-monotonic $j_c(B)$ dependence. We support that the secondary peak described by the Gaussian term is more suitable for the phase separation. Then $A$ determines the pinning increase due to rise of a non-superconducting inclusions, $B_w$ depends on a growth speed and a distribution of the inclusions size. The temperature evolution of the secondary peak on experimental magnetization loops may identify background process.

# 4 Conclusion

ECSM was modified to describe the secondary peak on magnetization loops. The same term is used to express the peak on the $j_c(B)$ dependence (7) and the depression on the $L_s(B)$ dependence (8). The height of the secondary peak on the $M^+(H)$ branch is determined by both $j_c(B)$ and $L_s(B)$ dependencies. Instead the peak on the $M^-(H)$ branch is influenced by the $j_c(B)$ dependence only. In the presented form (2-8) ECSM reproduces successfully the shape of magnetization loops at low and high magnetic fields.

**Acknowledgments** The work is supported by project № 20 of RAS Program "Quantum mesoscopic and disordered structures" and grant № 11-02-98007-р of Russian foundation for basic research.